\documentclass[prb,preprint]{revtex4-1} 

\usepackage{amsmath}  
\usepackage{amsfonts} 
\usepackage{graphicx}

\begin{document}

\title{An introduction to arrays of coupled waveguides}

\author{B. M. Rodr\'{\i}guez-Lara}
\email{bmlara@inaoep.mx} 
\affiliation{Instituto Nacional de Astrof\'{i}sica, \'{O}ptica y Electr\'{o}nica \\ Calle Luis Enrique Erro No. 1, Sta. Ma. Tonantzintla, Pue. CP 72840, M\'{e}xico} 

\author{Francisco Soto-Eguibar}
\affiliation{Instituto Nacional de Astrof\'{i}sica, \'{O}ptica y Electr\'{o}nica \\ Calle Luis Enrique Erro No. 1, Sta. Ma. Tonantzintla, Pue. CP 72840, M\'{e}xico}

\author{Demetrios N. Christodoulides}
\affiliation{CREOL/College of Optics and Photonics, University of Central Florida, Orlando, Florida, USA}

\date{\today}

\begin{abstract}
We present the theoretical basis needed to work in the field of photonic lattices. 
We start by studying the field modes in- and out-side a single waveguide. 
Then, we use perturbation theory to deal with an array of coupled waveguides and construct mode-coupling theory.
Finally, we realize a survey of optical analogues to quantum systems accompanied by an extensive bibliography.
\end{abstract}

\maketitle 

\section{Introduction}

It was 1974 when Tsai and Thomas proposed the analogy between an optical waveguide coupler and the quantum-mechanical double well in an American Journal of Physics letter.\cite{Tsai1974p636}
While the ultimate goal for the simulation of physics was set up by Feynman in 1985 in the form of quantum simulators and universal quantum computers,\cite{Feynman1982p467} classical optical analogues of quantum phenomena have proved an impressive tool to study otherwise theoretical physics in the laboratory.\cite{Campi1967p133}
In particular, arrays of more than two photonic waveguides have been a recursive example of classical simulators of quantum physics since the proposal of Christodoulides and Joseph to optically simulate nonlinear atomic chains.\cite{Christodoulides1988p794}
The coupling of waveguides is an old problem in optics\cite{Jones1965p261} and coupled mode theory is used to describe it.\cite{Snyder1972p1267,Kogelnik1972p2327,Yariv1973p919,Streifer1987p1,Haus1987p16,Huang1994p963} 
In the beginning, the first realizations of such photonic structures was just oriented toward their use in optical circuits\cite{Somekh1973p46} but modern fabrication techniques\cite{Davis1996p1729,Chan2001p1726} and paradigm shift\cite{Christodoulides1988p794,Christodoulides2003p817} has cemented their use as classical simulators of quantum systems, which is now helping back on the design of integrated photonic devices.\cite{ElGanainy2013p161105}

In the following we present a brief, basic tutorial including an extensive list of literature resources that aims to introduce undergrad students to the study of arrays of coupled waveguides. 
The theoretical core is presented in the two following sections where the concept of normal modes of a waveguide is introduced and, then, the basics of coupled mode theory are revisited through the introduction of perturbations in the equations of Maxwell. 
Later, we discuss the experimantal fabrication methods utilized to create these photonic structures and, finally, we explore the theory behind optical simulators of quantum physics with contemporary examples.

\section{Modes in single waveguides}

Waveguide structures support electromagnetic modes depending on their characteristics: geometry, dimensions, permitivity, permeability, etc. Let us start from Maxwell equations and solve them for an infinitely large, homogeneous, isotropic dielectric medium,
\begin{eqnarray}
\nabla \cdot \vec{D} &=& 0,  \label{eq:Maxwell1}\\
\nabla \cdot \vec{B} &=& 0,  \label{eq:Maxwell2}\\
\nabla \times \vec{E}&=& - \frac{\partial \vec{B}}{\partial t}, \label{eq:Maxwell3}\\
\nabla \times \vec{H}&=& \frac{\partial \vec{D}}{\partial t} ,  \label{eq:Maxwell4}
\end{eqnarray} 
that is invariant to translation over the $z$-axis. The electric field can be decomposed in transverse electric (TE) and magnetic (TM) components,\cite{Stratton1941,VolkeSepulveda2006p867} 
\begin{eqnarray} \label{eq:Efield}
\vec{E} &=& \left\{ c_{TE} \left( \hat{e}_{u} \frac{1}{h_{v}} \frac{d}{dv} + \hat{e}_{v} \frac{1}{h_{u}} \frac{d}{du} \right) \right. +  \nonumber \\
&& \left. c_{TM}  \left[ i k_{z} \left( \hat{e}_{u} \frac{1}{h_{u}} \frac{d}{du} + \hat{e}_{v} \frac{1}{h_{v}} \frac{d}{dv} \right) + \hat{e}_{z} k_{\perp}^{2}  \right] \right\} \psi(\vec{r}), 
\end{eqnarray}
where $c_{TE}$ and $c_{TM}$ are constants related to the transverse electric and transverse magnetic modes, respectively, the coordinate system $(u,v,z)$ is given by the unit vectors $\hat{e}_{j}$ and the scaling factors $h_{j}$, the wavevector $\vec{k}$ is given by its transverse, $k_{\perp}$, and axial, $k_{z}$, components such that $k^{2}= k_{\perp}^{2} + k_{z}^{2}$, and the Hertz potential answers the Helmholtz equation
\begin{eqnarray}
\left( \nabla^{2} + k^{2} \right)\psi(\vec{r})=0.
\end{eqnarray}
The magnetic field can be calculated from (\ref{eq:Maxwell4}) and the constitutive equations, $\vec{D} = \epsilon \vec{E}$ and $\vec{H} = \vec{B}/\mu$ with the permitivity and permeability of the material given by $\epsilon$ and $\mu$, in that order.

Now, we can assume a waveguide with polar symmetry in its transversal section; i.e., a circular-cylinder with radius $\rho=a$ and infinite length. The modes supported by this waveguide, $\rho < a$, are given by substituting in (\ref{eq:Efield}) the Hertz scalar potential
\begin{eqnarray}
\psi_{in}(\vec{r}) &=& J_{m}(k_{\perp} \rho) e^{i m \phi} e^{-i k_{z} z} e^{-i \omega t}, \label{eq:ScalarWaveIn}
\end{eqnarray}
where $J_{m}(x)$ are the Bessel functions of the first kind, with  $m=0,\pm1,\pm2, \ldots$. The transverse component of the wavevector is $k_{\perp} = \sqrt{ \omega^{2} \mu \epsilon - k_{z}^2}$ with $\mu$ and $\epsilon$ the permeability and permitivity of the materialm, respectively.
It is straightforward to show that the field modes defined by (\ref{eq:Efield}) and (\ref{eq:ScalarWaveIn}) form an orthogonal set that can be normalized if required.
We cannot forget the modes supported by the vacuum outside the waveguide core, $\rho > a$, but first let us think about a given guided mode. 
For a guided mode, the power of the electric field propagates through the waveguide; in other words, $k_{z} \sim k =\omega \sqrt{\mu \epsilon}$. 
This means that if the velocity of light in the waveguide core, $\sqrt{\mu \epsilon} $, is lower than in the exterior, in this case $\sqrt{\mu_{0} \epsilon_{0}}$, then $k_{z} > \omega \sqrt{\mu \epsilon_{0}} $ and $k_{\perp}$ will be purely imaginary.
Thus, we can write $\kappa_{\perp} = -i k_{\perp}$ and the scalar potential:
\begin{eqnarray}
\psi_{out}(\vec{r})&=& K_{m}(\kappa_{\perp} \rho) e^{i m \phi} e^{-i k_{z} z}  e^{-i \omega t},
\end{eqnarray}
where we are using the modified Bessel functions of the second kind, $K_{m}(x)$,
and we have to guarantee continuity of the whole electromagnetic field $\vec{E}_{in}(a)=\vec{E}_{out}(a)$, through an adequate choice of the constants $c_{TE}^{in}$, $c_{TM}^{in}$, $c_{TE}^{out}$ and $c_{TM}^{out}$.
Thus, we will have hybrid modes most of the time. The  asymptotic behavior of the modified Bessel function is $H_{m}(x) \propto e^{-i x}$ for $x\gg1$, which implies that in our case the field outside the waveguide core will exponentially decay with the distance from the boundary of the waveguide; i.e. it is an evanescent field.
So, we could handwave, that a second identical waveguide placed nearby, may harvest the electrical field in the first, thanks to this evanescent wave, coupling both waveguides; in what follows, we will revisit coupled mode theory formalism to show this.

\section{Mode coupling in arrays of waveguides}

Let us consider an array of waveguides and take one of them as the unperturbed waveguide and view the rest as perturbations.\cite{Snyder1972p1267,Kogelnik1972p2327,Yariv1973p919,Streifer1987p1,Haus1987p16,Huang1994p963} 
We can write the transversal forward propagating electric perturbed and unperturbed field modes at the $n$th waveguide as, 
\begin{eqnarray}
\vec{E}_{n,x} = a_{n,x}(z) \vec{e}_{n}(x,y)  e^{-i \omega t},  \quad x= p,u,
\end{eqnarray}
where the vector component $\vec{e}_{n}$ is one of the orthonormal transverse modes of the $n$th waveguide and the propagating field amplitude is such that for the unperturbed system $a_{n,u}(z) =  e^{i \beta_{n}(z)}$ and for the perturbed system $a_{n,p}(z) =  \mathcal{E}_{n}(z)$, we are gonna use this notation because we are gonna cycle through all the guides treating one at a time as unperturbed. 
We can introduce the perturbation into the equations of Maxwell through the polarization field $\vec{P}$ such that $\vec{D}= \epsilon \vec{E} + \vec{P}$ so
\begin{eqnarray}
\nabla \cdot \epsilon_{n} \vec{E}_{n} &=& 0,  \label{eq:MaxwellP1}\\
\nabla \cdot \vec{B}_{n} &=& 0,  \label{eq:MaxwellP2}\\
\nabla \times \vec{E}&=& i \omega \mu_{0} \vec{H}_{n}, \label{eq:MaxwellP3}\\
\nabla \times \vec{H}_{n} &=&- i \omega \left( \epsilon_{n} \vec{E}_{n} + \vec{P}_{n} \right), \label{eq:MaxwellP4}
\end{eqnarray} 
where the polarization field at the unperturbed waveguide is null, $\vec{P}_{n_{0}}=0$.
By use of all these definitions and some vector calculus identities, we can write Lorentz reciprocity between unperturbed and perturbed fields, 
\begin{eqnarray}
\int_{S} dS &&~ \nabla \cdot \left[ \vec{E}_{n,p} \times \vec{H}_{n,u}^{\ast} + \vec{E}_{n,u}^{\ast} \times \vec{H}_{n,p}  \right] \nonumber \\
&& = i \omega \int_{S} dS ~ \vec{P}_{n} \cdot \vec{E}_{n,u}^{\ast},  \label{eq:LorentzRec}
\end{eqnarray}
in the corresponding form
\begin{eqnarray}
\left( i \frac{d}{dz} + \beta(z) \right) && \mathcal{E}_{n}(z) \int_{S} dS ~ \left[ \vec{e}_{n} \times \vec{h}_{n,m}^{\ast} + \vec{e}_{n}^{\ast} \times \vec{h}_{n,m}  \right] \cdot \hat{e}_{z} \nonumber \\ 
&& = - \omega \int_{S} dS ~\vec{e}_{n_{0},m}^{\ast} \cdot \vec{P}_{n}. 
\end{eqnarray}
The integral on the left hand side in (\ref{eq:LorentzRec}) can be identified with the power of the field modes after we revisit their orthonormality,
\begin{eqnarray}
\int_{S} dS ~ \left( \vec{e}_{m} \times \vec{h}_{\tilde{m}}^{\ast} + \vec{e}_{\tilde{m}}^{\ast} \times \vec{h}_{m} \right) &=& 2 P_{0} ~\delta_{m,\tilde{m}}, \label{eq:LR2}
\end{eqnarray} 
where the constant $P_{0}$ is the normal power of the field mode and the symbol $\delta_{a,b}$ stands for the delta of Kronecker.
At this point we need to make an assumption to simplify our work.
We will assume that the differences between the waveguides are small, that is why we are treating them as perturbations.
If the differences are small, then the normal transversal modes in each waveguide are the same and we can write (\ref{eq:LR2}) in the form: 
\begin{eqnarray}
\left( i \frac{d}{dz} + \beta_{n} \right) a_{n}  = - \frac{\omega}{4 P_{0}} \int_{S} dS ~\vec{e}_{n}^{\ast} \cdot \vec{P}_{n}. \label{eq:PreDifSet}
\end{eqnarray}
Now, we need to find what the right hand side of the equation means.
We can divide the contributions to the polarization field in those coming from the local and coupling perturbations, 
\begin{eqnarray}
\vec{P}_{n} = \vec{P}_{n,l} + \vec{P}_{n,c}, \label{eq:Polarization}
\end{eqnarray}
with 
\begin{eqnarray}
\vec{P}_{n,l} &=& \Delta\epsilon_{n} \mathcal{E}(z) \vec{e}_{n}, \\
\vec{P}_{n,c} &=& \sum_{j=0}^{N-1} \Delta\epsilon_{n} a_{j}(z) \vec{e}_{j},
\end{eqnarray}
where the functions $\Delta\epsilon_{n}$ are the deviations from the unperturbed permitivity $\epsilon$.
Note that we call $\vec{P}_{n,c}$ the contribution to the polarization field from the coupling due to it accounting for all the other waveguides in the system.
Then, we can define two contributions by substituting (\ref{eq:Polarization}) into the left hand side of (\ref{eq:PreDifSet}),
\begin{eqnarray}
\alpha_{n} &=& \frac{\omega}{2 P_{0}} \int_{S}dS~ \Delta\epsilon_{n} ~ \vec{e}_{n}^{\ast}\cdot\vec{e}_{n}, \\
c_{n,j} &=& \frac{\omega}{4 P_{0}} \int_{S}dS~ \Delta\epsilon_{n} ~ \vec{e}_{n}^{\ast}\cdot\vec{e}_{j}.
\end{eqnarray}
The first comes from the local polarization field and those terms with $j=n$ from the coupling polarization field. 
The other comes from the rest of the terms,  $j \ne n$, of the coupling polarization field. 
Then we can describe the field dynamics in our array of coupled waveguides as the differential equation set,
\begin{eqnarray}
\left( i \frac{d}{dz} + \beta_{n} + \alpha_{n} \right) \mathcal{E}_{n} + \sum_{j=0, j\ne n}^{N-1} c_{n,j} \mathcal{E}_{j}  = 0. \label{eq:GenEq}
\end{eqnarray}
If we are just interested in nearest neighbor coupling, then we can rewrite the differential set above in the form: 
\begin{eqnarray}
-i \frac{d}{dz} \mathcal{E}_{0} &=&  n_{0}(z) \mathcal{E}_{0} +  g_{1}(z)\mathcal{E}_{1}, \\
-i \frac{d}{dz} \mathcal{E}_{1} &=&  n_{1}(z) \mathcal{E}_{1} +  g_{1}(z)\mathcal{E}_{0} +  g_{2}(z)\mathcal{E}_{2}, \\
\vdots && \\
-i \frac{d}{dz} \mathcal{E}_{N-2} &=&  n_{N-2}(z) \mathcal{E}_{N-2} +  g_{N-2}(z)\mathcal{E}_{N-3} + \nonumber \\
&& +  g_{N-1}(z)\mathcal{E}_{N-1}, \\
-i \frac{d}{dz} \mathcal{E}_{N-1} &=&  n_{N-1}(z) \mathcal{E}_{N-1} +  g_{N-1}(z)\mathcal{E}_{N-2}. \label{eq:DifSet}
\end{eqnarray}
We will refer to the parameter $n_{j}(z)$ as effective refractive index due to its relation to the permeability and its perturbations in the $j$th waveguide and to the parameter $g_{j}(z)$ as first neighbor couplings.
If we need to  deal with $k$th neighbor couplings, we can do it starting from (\ref{eq:GenEq}).

\section{Experimental fabrication}

The first array of optical waveguides reported in the literature was fabricated by proton implantation in GaAs\cite{Somekh1973p46,Garvin1973p455}
Later, the development of high-energy femtosecond pulse lasers prompted the study of photochemical reactions in different glasses for their used in integrated optical circuits.
In short, when a high-intensity ultra-short laser pulse is focused inside a material, absoption occurs and a microplasma is formed due to optical breakdown which induces permanent changes in the refractive index of the damaged area.\cite{Davis1996p1729}
These changes are highly reproducible and, by moving the sample or the focused laser,
waveguides can be written in bulk materials; e.g. fused and doped silica.\cite{Davis1996p1729,Miura1997p3329,Glezer1997p882,Chan2001p1726,Szameit2010p163001}
Typically, a high-energy pulsed Ti:Sapphire laser is used to process the glass material; e.g.  800nm with power in the hundreds of mW for fused silica. 
The laser light is focused through an optical system that scans and damages the material at a given rate and can create low-loss three-dimensional photonic structures. 
The process has also been shown to reduce the nonlinear refractive index of written waveguides.\cite{Blomer2006p2151,Fleischer2005p1780,Christodoulides1988p794}
It is also possible to create waveguide arrays susceptible of real-time modifications by use of thermo-optic materials.\cite{Pertsch2002p3247}

\section{Optical analogues}

Notice that if we complex conjugate \eqref{eq:DifSet}, it can be written in the short form
\begin{eqnarray}
\frac{d}{dz} \mathbf{E}^{\ast} = -i \mathbf{H} \mathbf{E}^{\ast}, \label{eq:ShortForm}
\end{eqnarray}
where the asterisk stands for complex conjugation and $\mathbf{H}$ is a real symmetric tridiagonal matrix.
Thus, it has the form of a Schr\"odinger equation where the conjugated field amplitudes play the role of the wavefunction coefficients and the propagation variable that of time.
This allows us to make classical optics analogues of quantum mechanical systems both non-relativistic and non-relativistic.

\subsection{Displaced number states}

Let us take as our first example the so-called Glauber-Fock photonic lattice\cite{PerezLeija2010p2409,PerezLeija2011p1833,Keil2011p103601,RodriguezLara2011p053845,Keil2012p3801,PerezLeija2012p013848}  
\begin{eqnarray}
 -\imath \frac{d \mathcal{E}_{0}(z)}{d z} &=&   \mathcal{E}_{1}(z), \\
 -\imath \frac{d \mathcal{E}_{j}(z)}{d z} &=&  \sqrt{j} \mathcal{E}_{j+1}(z) + \sqrt{j-1} \mathcal{E}_{j-1}(z), \\
 -\imath \frac{d \mathcal{E}_{N-1}(z)}{d z} &=& \sqrt{N-2} \mathcal{E}_{N-2}(z),
\end{eqnarray}
where we can the creation and annihilation operators, $\hat{a}$ and $\hat{a}^{\dagger}$, which in matrix form have elements $a_{m,n}= \sqrt{m} \delta_{m+1,m}$ and  $a_{m,n}^{\dagger}= \sqrt{m}  \delta_{m,m+1}$ with $m=0, 1, \ldots$, and rewrite in the following form:
\begin{eqnarray}
\frac{d \mathbf{E}(z)}{d z} = i \left( \hat{a}^{\dagger} + \hat{a} \right)  \mathbf{E}(z), 
\end{eqnarray}
which, after complex conjugation, is identical to \eqref{eq:ShortForm} and accepts a solution of the form $\mathbf{E}(z) = e^{i z  \left( \hat{a}^{\dagger} + \hat{a} \right) } \mathbf{E}(0)$ that can be seen as a displaced superposition of Fock states; e.g. in the case of just $\mathcal{E}_{0}=1$ and all other field amplitudes equal to zero, the amplitude of the electric field at the $n$th waveguide will be equivalent to the complex conjugat of the $n$th amplitude of a coherent state with coherent parameter $iz$, $\mathcal{E}_{n}(z) = \langle n \vert z \rangle = e^{-z^2/2} (iz)^{n} / \sqrt{n!}$.
One could also treat the case of squeezed states, 
\begin{eqnarray}
\frac{d \mathbf{E}(z)}{d z} = i \left( \hat{a}^{\dagger 2} + \hat{a}^{2} \right)  \mathbf{E}(z), 
\end{eqnarray}
by means of two uncoupled photonic lattices\cite{Nezhad2013p023801} and extend it to $j$ uncoupled photonic lattices for couplings of the form $\left( \hat{a}^{\dagger j} + \hat{a}^{j} \right)$.

Recently, it has been shown that the creation of other classical analogues of non-classical states, such as W-states\cite{Rai2009p053849,PerezLeija2013p013842} or squeezed states,\cite{Sukhorukov2013p053823} is possible in arrays of one- and two-dimensional waveguides.
Furthermore, by using intensity correlation schemes like the Hanbury Brown-Twiss correlations it is possible to study the path-entanglement created by propagation of photonic lattices.\cite{Bromberg2009p253904,Keil2010p023834}

\subsection{A two-level atom coupled to a quantum field}

The Jaynes-Cummings model is a work horse in quantum optics and describes systems in cavity-, trapped-ion- and circuit-QED (quantum electrodynamics) but the Rabi model may be realized only in circuit-QED and, then, the inclusion of nonlinear processes may be out of experimental reach at the time.
Arrays of coupled waveguides allow the classical simulation of a general nonlinear Rabi model\cite{RodriguezLara2013p12888}
\begin{eqnarray} \label{eq:OurH}
\hat{H} &=& h(\hat{n}) + \frac{\omega_{0}}{2} \hat{\sigma}_{z} + g_{-} \left( \hat{a} \frac{f(\hat{n})}{\sqrt{\hat{n}}} \hat{\sigma}_{+} +  \frac{f(\hat{n})}{\sqrt{\hat{n}}} \hat{a}^{\dagger} \hat{\sigma}_{-}   \right) + \nonumber \\
&&  g_{+} \left( \hat{a} \frac{f(\hat{n})}{\sqrt{\hat{n}}} \hat{\sigma}_{-} +  \frac{f(\hat{n})}{\sqrt{\hat{n}}} \hat{a}^{\dagger} \hat{\sigma}_{+}   \right), \label{eq:NonLinRabi}
\end{eqnarray}
where the real functions $f(\hat{n})$, $g(\hat{n})$ and $h(\hat{n})$ can be nonlinear functions of the number operator, $\hat{n}= \hat{a}^{\dagger} \hat{a}$, and the parameters $\omega_{0}$ and $g_{\pm}$ are the qubit frequency, rotating and counter-rotating couplings, in that order.
The model describes a pletora of qubit-field interactions\cite{Longhi2011p3407,Crespi2012p163601,PerezLeija2012p014023,RodriguezLara2013p87,RodriguezLara2013p12888} and can be easily set in the short form (\ref{eq:ShortForm}) by using a parity decomposition, $\vert \psi_{\pm} \rangle =  \sum_{j} \mathcal{E}^{\pm}_{j}(z) \vert \pm,j \rangle$ with $\vert +(-),j \rangle = \hat{\sigma}_{x}^{j} \vert g(e), j \rangle$, and the matrix representation of Pauli matrices and bosonic creation and annihilation operators used before,\cite{RodriguezLara2013p12888}
\begin{eqnarray}
i \frac{d}{d z} \mathcal{E}_{0}^{(\pm)}  &=&  d^{(\pm)}(0) \mathcal{E}_{0}^{(\pm)} + g_{\pm} f(1) \mathcal{E}_{1}^{(\pm)}, \\
i \frac{d}{d z} \mathcal{E}^{(\pm)}_{2k+1} &=& d^{(\pm)}(2k+1) \mathcal{E}_{2k+1}^{(\pm)} + g_{\pm} f(2k+1) \mathcal{E}_{2k}^{(\pm)} + \nonumber \\
&&   g_{\mp} f(2k+2) \mathcal{E}_{2k+2}^{(\pm)}, \quad k \ge 0,\\
i \frac{d}{d z} \mathcal{E}^{(\pm)}_{2k} &=& d^{(\pm)}(2k) \mathcal{E}_{2k}^{(\pm)} + g_{\mp} f(2k) \mathcal{E}_{2k-1}^{(\pm)} + \nonumber \\
&& g_{\pm} f(2k+1) \mathcal{E}_{2k+1}^{(\pm)}, \qquad \qquad k \ge 1,
\end{eqnarray}
with $d^{(\pm)}(j) = h(j) \mp (-1)^{j} \frac{\omega_{0}}{2}$.
Thus, we end up with two one-dimensional arrays of coupled waveguides; each one describing a parity subspace of the qubit-field system.

\subsection{Three-level atoms driven by classical fields}

The differential set for three non-identical waveguides coupled between them,
\begin{eqnarray}
\frac{d \mathbf{E}^{\ast}(z)}{d z} = -i \left(
\begin{array}{ccc}
n_{1}(z) & \alpha(z) & 0 \\
\alpha(z) & n_{2}(z) & \beta(z) \\
0 & \beta(z) & n_{3}(z)
\end{array}
\right) \mathbf{E}^{\ast}(z), 
\end{eqnarray}
is equivalent to the dynamics of the population amplitudes of a class of three-level atom driven by two classical fields, where the refractive index are equivalent to the role of the self-energies of the atomic states and the couplings are the analogue of time-dependent classical driving fields coupling the atomic states.
This allows us to use all the machinery developed in quantum optics to deal with these optical systems; e.g. construct propagators based on the Lie group generators of $su(3)$.\cite{Dattoli1991p1247}

One can recognize valuable quantum optical phenomena and use it for photonic circuitry purposes; e.g. one-directional adiabatic three-waveguide couplers \cite{Milton1975p1207,Schneider2001p129,Huang2006p056606,Paspalakis2006p30,Longhi2007p201101,Lahini2008p193901,DellaValle2008p011106,Salandrino2009p4524,Longhi2009p243} that can be considered the optical analogue of stimulated Raman adiabatic passage (STIRAP). \cite{Bergmann1998p1003}
As an example, we can set $n_{j}=0$, $\alpha(z) = c e^{-(z-\zeta)^2 / \zeta^2 }$ and $\beta(z) = c e^{-z^2 / \zeta^2 }$ where $c, \zeta \in \mathbb{R}$\cite{Mitra2003p043409} and obtain perfect transfer from the first to the last waveguide.

It is also possible to identify that the parameter set  $ \alpha(z) = \beta(z) = c$, $\gamma(z) = - m \cos(z) / 2$ and $\delta(z) = \sqrt{3} \gamma(z)$ with $c,m \in \mathbb{R}$ provides the so-called atomic population trapping\cite{Harshawardhan1997p2165} whenever the value of the constant $m$ produces a zero in the zeroth order Bessel function, $J_{0}(m)=0$. 
Thus, we can use an optical analogue to produce coherent oscillations of the field amplitudes through the photonic device that deliver intensity trapping at certain intervals for a finite propagation distance.\cite{RodriguezLara2013}

\subsection{Majorana \& Dirac physics}

The Majorana equation \cite{Majorana1932p335} is the first construction using the infinite dimensional representation of the Lorentz group to present a relativistically invariant theory of arbitrary half integer or integer spin particles that provides a mass spectrum for elementary particles \cite{Fradkin1966p314},
\begin{eqnarray} \label{eq:MajoranaEquation}
\imath \hbar\gamma^\mu \partial_\mu \psi = m\psi_c,
\end{eqnarray}
where Dirac representation has been used; the symbol $\psi_c$ stands for charge conjugation of the spinor $\psi$, $\psi_{c} \equiv \gamma^2 \psi$, and $\gamma_{\mu}$ are the Dirac matrices.
In (1+1) dimensions, it can be reduced to two Dirac equations for Majorana fermions with opposite mass signs,\cite{Noh2013p040102}
\begin{eqnarray}
\imath \hbar \partial_{t} \phi_{\pm} = \left( c \hat{p}_{x} \hat{\sigma}_{x} \pm m c^2 \hat{\sigma}_{z}  \right) \phi_{\pm},
\end{eqnarray}
for the Majorana fermions $ \phi_{\pm} \in \mathbb{C}^2$ , \textit{i.e.}, $ \phi_{\pm} = \hat{\sigma}_{z} \hat{\sigma}_{x} \phi_{\pm}^{\ast}$.
Here we can borrow the definition of the linear momentum operator in terms of the creation/anihilation operators, $\hat{p}_{x}= (\hat{a}^{\dagger} + \hat{a})/\sqrt{2}$, and realize that it is identical to \eqref{eq:NonLinRabi} by setting $\hbar=1$, $h(\hat{n})=0$, $\omega_{0} = \pm 2 mc^{2}$, $f(\hat{n}) = \sqrt{n}$ and $g_{+} = g_{-} = c / \sqrt{2}$.\cite{RodriguezLara2013p13101400}
Thus, we can classically simulate a Majorana equation via four one-dimentional photonic crystals where we can observe a variety of relativistic phenomena.
The Dirac equation has already been studied in the context of binary matrices,\cite{Longhi2011p453,Longhi2010p022118,RodriguezLara2013p038116,Tran2014p179} e.g., zitterbewegun in the form of oscillations in the baricenter of the field amplitude as it propagates\cite{Longhi2010p235} or Klein tunneling as a reflection of the baricenter at the interface between two binary lattices.\cite{Dreisow2010p143902,Longhi2010p075102,Dreisow2012p10008}

\subsection{Quantum billiards and walks}

Imagine the simplest quantum walk, the one where at each step you have half-half chance of walking left or right and the probability to find the walker here or there is given by quantum interference of the paths.
That is equivalent to an electromagnetic field propagating through a one-dimensional array of identical waveguides homogeneously coupled by nearest neighbor interactions.\cite{Jones1965p261,Somekh1973p46,Perets2008p170506,Longhi2008p035802}
The variation of the coupling parameters affects directly the left/right walking probability. 
The introduction of different refractive indices gain at the waveguides, the use of non-classical light states, or all at the same time allows for the simulation of nonlinear quantum random walks.\cite{Rai2008p042304}
The analogue can be extended to two-dimensions and, then, we can speak of classical simulation of quantum billiards.\cite{Krimer2011p041807(R)}
It will come a time when multiple single photons propagating through optical waveguide arrays will be used to generate proper quantum random walks with high-order correlation functions \cite{Gard2013p1538}.

\subsection{Electron propagation in strucures}

Photonic lattices provide us with a classical simulator for tight-binding models of a particle in potentials such as an electron in crystals or quasi-crystals as the quantum dynamics for such a system, provided by the Hamiltonian
\begin{eqnarray}
\hat{H}= \sum_{j} w_{j} \hat{c}_{j}^{\dagger} \hat{c}_{j} - \sum_{j} t_{j} \left( \hat{c}_{j}^{\dagger} \hat{c}_{j+1} + \hat{c}_{j+1}^{\dagger} \hat{c}_{j} \right),
\end{eqnarray}
where $\hat{c}_{j}$ ($\hat{c}_{j}^{\dagger}$)is the particle annihilation (creation) operator at site $j$, reduces to that of a classical field propagating in a photonic lattice for a single electron system.
Thus, we can use waveguide arrays to study a range of physical phenomena, e.g. Bloch oscillations,\cite{Longhi2011p3248,Longhi2012p075143,Gordon2004p2752} Bloch-Zenner oscillations, \cite{Longhi2006p416,Dreisow2009p076802,Zheng2010p3865,Longhi2012p075144,Arevalo2011p60011,Zheng2011p1339,Zheng2010p3865} coherent tunnelling,\cite{Tsai1974p636,Trompeter2006p023901} and Zeno effect\cite{Longhi2006p110402,Dreisow2008p143602,Biagioni2008p3762,Longhi2009p243} to mention a few.
In particular, electrons propagating in weakly disordered solids are predicted to localize. 
This absence of diffusion in certain random lattices known as Anderson localization can be classically simulated in batches of static or real-time thermally induced photonic lattices with controlled disorder.\cite{Schwartz2007p52,Longhi2009p243,Lahini2010p163905,Thompson2010p053805,Martin2011p13636,Abouraddy2012p040302,Dufour2012p155306,ElGanainy2012p64004,Naether2012p485,Naether2012p485,Giuseppe2013p150503,Segev2013p197,Karbasi2013p1452}
One can also simulate the effect of non-classical correlations on the localization of separate electrons with light propagation through disordered lattices.\cite{Lahini2011p041806(R),Giuseppe2013p150503}

By engineering higher dimensional photonic structures, one could simulate a larger family of Hamiltonians involving an electron in the presence of structured potentials, 
\begin{eqnarray}
\hat{H}= \sum_{j} w_{j} \hat{c}_{j}^{\dagger} \hat{c}_{j} - \sum_{(j,k)} t_{j,k}  \hat{c}_{j}^{\dagger} \hat{c}_{k},
\end{eqnarray}
and study optical analogues of phenomena such as dynamic localization in the presence of periodic potentials and bias fields.\cite{Szameit2009p271,Szameit2010p223903}
The use of honeycomb lattices with helical modulation allows for the simulation of graphene physics; e.g. band collapse,\cite{Crespi2013p013012} and topological insulators.\cite{Rechtsman2013p196,Rechtsman2012,Rechtsman2012b}

\section{Conclusions}

Arrays of coupled waveguides have been used in a multitude of contexts in the literature. 
They have been designed to be integrated in optical circuits as waveguide couplers\cite{Milton1975p1207,Schneider2001p129,Huang2006p056606,Paspalakis2006p30,Longhi2007p201101,Lahini2008p193901,DellaValle2008p011106,Salandrino2009p4524,Longhi2009p243,Bergmann1998p1003,RodriguezLara2013}, multiplexors\cite{Rai2009p053849,PerezLeija2013p013842}, light rectifiers\cite{Longhi2009p458} and on demand sources of tailored guided modes\cite{Trompeter2003p3404}.
It is possible to demonstrate diffraction phenomena such as multiband diffraction and refraction,\cite{Longhi2006p1857} generation of surface waves\cite{Garanovich2008p203904,Szameit2008p203902,Malkova2009p1633,Malkova2009p043806} and  light bullets in nonlinear periodically curved waveguides,\cite{Matuszewski2010p043833} the acceleration of Wannier-Stark states in uniform optical lattices,\cite{ElGanainy2011p023842} generation of two-dimensional Airy-like\cite{Deng2013p1404} and other structures\cite{Garanovich2012,PerezLeija2013p17951} with the goal of controlling light propagation.\cite{Hizanidis2008p18296,Chacon2012p013813,Cao2012p19119}
Some structures can provide self-focusing,\cite{Christodoulides1988p794} others perfect imaging\cite{Keil2012p809} and some others transport control through the phase of the impinging fields.\cite{Thompson2011p214302}
It also is possible to use supersymmetric quantum mechanics to design isospectral one-dimensional crystals\cite{Miri2013p233902,Longhi2013p40008,ZunigaSegundo2013} or produce resonant propagation via local $\mathcal{PT}$ invariance.\cite{ElGanainy2012p033813}
Two-dimensional structures can simulate effective electromagnetc fields for bosons. \cite{Longhi2012p042104,Longhi2013p3570}
The use of waveguide arrays combining gain and loss allows us to classically explore the Hubbard model\cite{Longhi2011p155101}, highly-correlated states\cite{Longhi2011p033821} and non-Hermitian $\mathcal{PT}$-symmetric systems.\cite{Longhi2009p123601,Longhi2010p032111,Longhi2012p012112,Ramezani2012p013818,Joglekar2011p063817,Vemuri2013p044101,Vemuri2011p043826}
There will come a point when these photonic structures will be integrated with single-photon sources for quantum state transport.\cite{PerezLeija2013p012309,Perez-Leija2013p022303}
The field is open and the literature increasing day by day, today is a good day to join it.

\end{document}